\documentclass[submission,copyright,creativecommons,noderivs,noncommercial]{eptcs}
\providecommand{\event}{FMAS 2024} 

\usepackage{times}
\usepackage{amssymb}
\usepackage{amsmath}

\usepackage{bm}
\usepackage{bbm}

\usepackage{bsymb}

\usepackage[mathscr]{eucal}
\usepackage{eufrak}
\usepackage{pifont}

\usepackage{caption}
\usepackage{subcaption}

\usepackage[T1]{fontenc}
\usepackage{graphicx}
\usepackage{comment}
\usepackage{url}

\usepackage{wrapfig}

\usepackage{afterpage}

\usepackage[usenames,dvipsnames]{xcolor}

\newcommand{\veryslightlylongpage}{\enlargethispage{0.2\baselineskip}}

\allowdisplaybreaks

\usepackage{iftex}

\ifpdf
  \usepackage{underscore}         
  \usepackage[T1]{fontenc}        
\else
  \usepackage{breakurl}           
\fi

\title{Autonomous System Safety Properties\\
with Multi-Machine Hybrid Event-B}
\author{Richard Banach
\institute{Department of Computer Science,\\
University of Manchester, Manchester, M13 9PL, UK}
\email{richard.banach@manchester.ac.uk}
}
\def\titlerunning{Autonomous System Safety with Multi-Machine Hybrid Event-B}
\def\authorrunning{R. Banach}
\begin{document}
\maketitle

\begin{abstract}
Event-B is a well known methodology for the verified design and development
of systems that can be characterised as discrete transition systems. Hybrid
Event-B is a conservative extension that interleaves the discrete transitions
of Event-B (assumed to be temporally isolated) with episodes of continuously
varying state change. While a single Hybrid Event-B machine is sufficient for
applications with a single locus of control, it will not do for autonomous
systems, which have several loci of control by default. Multi-machine Hybrid
Event-B is designed to allow the specification of systems with several loci
of control. The formalism is succinctly surveyed, pointing out the subtle
semantic issues involved. The multi-machine formalism is then used to specify
a relatively simple incident response system, involving a controller, two
drones and three responders, working in a partly coordinated and partly
independent fashion to manage a putative hazardous scenario.
\end{abstract}

\section{Introduction}

These days, there is an inexorable drive to automate as many of the
tasks that humans engage in to progress their everyday lives as possible.
The motives range from gaining efficiency, to enabling genuinely new
activities which would otherwise be impossible. Autonomous systems can
be deployed to achieve both aims. Properly designed, not only can they
enable efficiencies in activities traditionally undertaken by humans
(e.g.~logistics), but they can also enable activities (e.g.~reactor
core investigation) that would be lethal for humans.

Proper design of autonomous systems implies the ability to place great
trust in their operation, given that they act in a manner much less
supervised than is the case for conventional systems. If a system
is to be given responsibility for deciding its own course of action (to
whatever degree is considered reasonable), it must be understood how the
gamut of its decision making capabilities keep it aligned with the widest
considerations that its behaviour may affect. In other words, the boundary
between the autonomous system and the wider environment needs to be reliably
appreciated, and the greater the creativity that the system is permitted
to display in meeting its local challenges, the greater the burden of
understanding that is imposed on system design to ensure that letting
the system do its thing will not jeopardise either itself, or elements
of the wider environment.

To try to ensure this, all available techniques for enhancing system
assurance may be brought to bear. As well as the traditional approaches
of careful design and testing, more formal approaches, that aim to
strengthen the guarantees that can be given regarding system behaviour,
may be used. This highlights safety properties, the focus of the B-Method.

The significant self-agency of autonomous systems means that they are
often cyber-physical systems \cite{Car.Rob...:06,Tab:09,GeiBro:15}.
The consequent interaction between discrete and continuous dynamics
raises a challenge for verification \cite{Garo:19,Sanf:21}. The
contribution of this paper is to show that the architecture and
capabilities of multi-machine Hybrid Event-B are suitable for
formalising the challenge just described.

Hybrid Event-B was envisaged as a conservative extension of conventional
Event-B that was able to accommodate continuous and smooth behaviour in
a rigorous way. The foundations of the formalism were investigated in
three papers \cite{BaHyEvB-I:15a,BaHyEvB-TWO:15b,BaHyEvB-3:15c},
referred to below as PaperI, PaperII, PaperIII. These may be consulted
for full technical details; a less detailed overview appears in
\cite{BaHyEvB-OV:24}. Of particular relevance to the present work is
\cite{BaHyEvB-TWO:15b}, in which the multi-machine version of the
formalism was explored, arriving at a framework that copes equally
comfortably with multi-machine continuous behaviour as it does with
the more familiar discrete behviours more typically used in verification.
Given that autonomous systems perforce exhibit a degree of disconnect
between themselves and any environment, a modelling and verification
framework that captures this structural aspect in separate constructs
is intrinsically useful. The main thrust of the present paper is the
assertion that these aspects of the multi-machine version of Hybrid
Event-B are particularly expressive and helpful in organising the
exploration and verification of the complex scenarios that arise when
autonomous systems are investigated.

The rest of the paper is as follows. Section \ref{sec-eb+heb} introduces
Hybrid Event-B as a natural extension of Event-B. Section \ref{sec-heb-ref} introduces Hybrid Event-B refinement. Section \ref{sec-heb-mult} presents
the main features of multiple cooperating Hybrid Event-B machines. Section
\ref{sec-inc-resp} applies the multi-machine formalism to present a simple
incident response system, involving a controller, two drones and three
responders, working in a partly coordinated and partly independent fashion
to manage a putative hazardous scenario. Each of the agents mentioned is
specified in its own machine, and the architecture of the multi-machine
formalism conveniently captured the cooperation mechanisms needed. Section
\ref{sec-verif} briefly considers verification for Hybrid Event-B and of
the case study, and Section \ref{sec-conc} concludes.

\section{From Event-B to Hybrid Event-B}
\label{sec-eb+heb}

\begin{figure}[b]
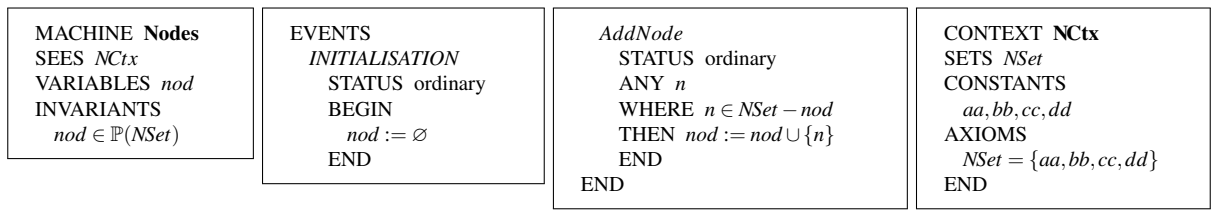

\fbox{\parbox[t]{0.19\linewidth}{\scriptsize 
\hspace*{2ex}MACHINE~~$\bf Nodes$\rule{0pt}{2.5ex}\\
\hspace*{2ex}SEES~~$N\!Ctx$\\
\hspace*{2ex}VARIABLES~~$nod$\\
\hspace*{2ex}INVARIANTS\\
\hspace*{4ex}$nod \in \mathbb{P}(N\!Set)$
\raisebox{-1ex}{\rule{0pt}{2.5ex}}
}}
\hfill
\fbox{\parbox[t]{0.22\linewidth}{\scriptsize 
\hspace*{2ex}EVENTS\rule{0pt}{2.5ex}\\
\hspace*{4ex}$\mathit{INITIALISATION}$\\
\hspace*{6ex}STATUS~~ordinary\\
\hspace*{6ex}BEGIN\\
\hspace*{8ex}$nod := \emptyset$\\
\hspace*{6ex}END
\raisebox{-1ex}{\rule{0pt}{2.5ex}}
}}
\hfill
\fbox{\parbox[t]{0.28\linewidth}{\scriptsize 
\hspace*{4ex}$AddNode$\rule{0pt}{2.5ex}\\
\hspace*{6ex}STATUS~~ordinary\\
\hspace*{6ex}ANY~~$n$\\
\hspace*{6ex}WHERE~~$n \in N\!Set - nod$\\
\hspace*{6ex}THEN~~$nod := nod \cup \{n\}$\\
\hspace*{6ex}END\\
\hspace*{2ex}END
\raisebox{-1ex}{\rule{0pt}{2.5ex}}
}}
\hfill
\fbox{\parbox[t]{0.23\linewidth}{\scriptsize 
\hspace*{2ex}CONTEXT~~$\bf N\!Ctx$\rule{0pt}{2.5ex}\\
\hspace*{2ex}SETS~~$N\!Set$\\
\hspace*{2ex}CONSTANTS\\
\hspace*{4ex}$aa , bb , cc , dd$\\
\hspace*{2ex}AXIOMS\\
\hspace*{4ex}$N\!Set = \{aa , bb , cc , dd\}$\\
\hspace*{2ex}END \raisebox{-1ex}{\rule{0pt}{2.5ex}}
}}
\caption{A simple Event-B machine, together with its context.}
\label{fig-eb-mch}
\end{figure}

The B-Method \cite{Bbook:1996} and Event-B in particular \cite{Ebook:2010}
have, by now, a well established pedigree. By way of introduction,
Fig.~\ref{fig-eb-mch} shows a simple Event-B machine $\bf Nodes$.
A static node set $N\!Set$ is defined in the context $\bf N\!Ctx$, and
a dynamic set $nod$ can add elements of $N\!Set$ to itself via the guarded
event $AddNode$. The dynamics of the machine consists of the state
transitions that ensue as events are executed one by one. Each possible
sequence of transitions can be collected into a system trace. It is assumed
that, in the real world, some non-empty interval of real world time elapses
between each event occurrence.

\begin{figure}
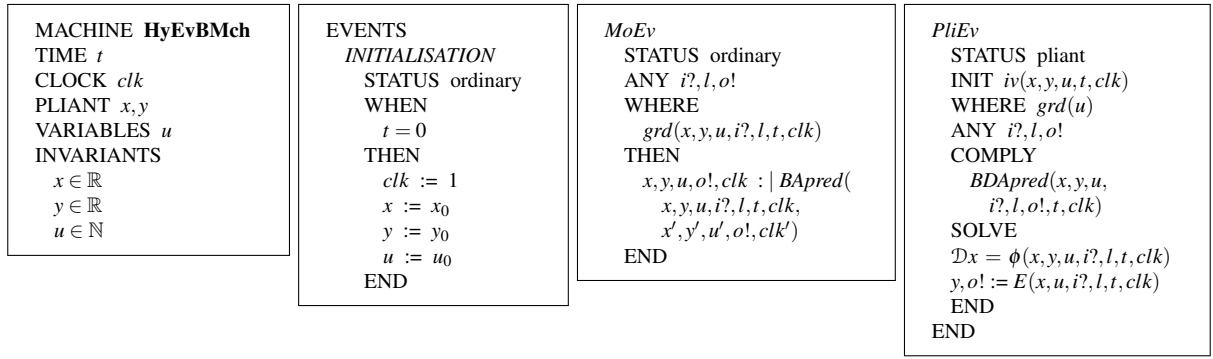

\fbox{\parbox[t]{0.22\linewidth}{\scriptsize 
\hspace*{2ex}MACHINE~~$\bf HyEvBMch$\rule{0pt}{2.5ex}\\
\hspace*{2ex}TIME~~$t$\\
\hspace*{2ex}CLOCK~~$clk$\\
\hspace*{2ex}PLIANT~~$x,y$\\
\hspace*{2ex}VARIABLES~~$u$\\
\hspace*{2ex}INVARIANTS\\
\hspace*{4ex}$x \in \mathbb{R}$\\
\hspace*{4ex}$y \in \mathbb{R}$\\
\hspace*{4ex}$u \in \mathbb{N}$
\raisebox{-1ex}{\rule{0pt}{2.5ex}}
}}
\hfill
\fbox{\parbox[t]{0.21\linewidth}{\scriptsize 
\hspace*{2ex}EVENTS\rule{0pt}{2.5ex}\\
\hspace*{4ex}$\mathit{INITIALISATION}$\\
\hspace*{6ex}STATUS~~ordinary\\
\hspace*{6ex}WHEN\\
\hspace*{8ex}$t = 0$\\
\hspace*{6ex}THEN\\
\hspace*{8ex}$clk$ ~:=~ $1$\\
\hspace*{8ex}$x$ ~:=~ $x_0$\\
\hspace*{8ex}$y$ ~:=~ $y_0$\\
\hspace*{8ex}$u$ ~:=~ $u_0$\\
\hspace*{6ex}END
\raisebox{-1ex}{\rule{0pt}{2.5ex}}
}}
\hfill
\fbox{\parbox[t]{0.25\linewidth}{\scriptsize 
\hspace*{2ex}$\mathit{MoEv}$\rule{0ex}{2.5ex}\\
\hspace*{4ex}STATUS~~ordinary\\
\hspace*{4ex}ANY~~$i?,l,o!$\\
\hspace*{4ex}WHERE\\
\hspace*{6ex}$grd(x,y,u,i?,l,t,clk)$\\
\hspace*{4ex}THEN\\
\hspace*{6ex}$x,y,u,o!,clk ~:|~ \mathit{BApred}($\\
\hspace*{8ex}$x,y,u,i?,l,t,clk,$\\
\hspace*{8ex}$x',y',u',o!,clk')$\\
\hspace*{4ex}END
\raisebox{-1ex}{\rule{0pt}{2.5ex}}
}}
\hfill
\fbox{\parbox[t]{0.24\linewidth}{\scriptsize 
\hspace*{2ex}$\mathit{PliEv}$\rule{0ex}{2.5ex}\\
\hspace*{4ex}STATUS~~pliant\\
\hspace*{4ex}INIT~~$iv(x,y,u,t,clk)$\\
\hspace*{4ex}WHERE~~$grd(u)$\\
\hspace*{4ex}ANY~~$i?,l,o!$\\
\hspace*{4ex}COMPLY\\
\hspace*{6ex}$\mathit{BDApred}(x,y,u,$\\
\hspace*{8ex}$i?,l,o!,t,clk)$\\
\hspace*{4ex}SOLVE\\
\hspace*{4ex}$\mathscr{D}x\,=\,\phi(x,y,u,i?,l,t,clk)$\\
\hspace*{4ex}$y,o!$ := $E(x,u,i?,l,t,clk)$\\
\hspace*{4ex}END\\
\hspace*{2ex}END
\raisebox{-1ex}{\rule{0pt}{2.5ex}}
}}
\caption{A schematic Hybrid Event-B machine.}
\label{fig-heb-mch}
\end{figure}

The restriction to discrete events makes the original Event-B poorly
suited for investigating the hybrid and cyber-physical systems that
are prevalent today. But the real-time gap assumed to be present
between discrete event occurrence, makes a convenient opening for
interleaving pieces of continuous behaviour into the time intervals
between them, and this was how Hybrid Event-B was designed.

The intuition just described results in there being two kinds
of event in Hybrid Event-B. There are {\bf mode} events, which
specify discrete changes of state ---just as in Event-B--- and there
are {\bf pliant} events, which specify episodes of continuous
state update. Since we have continuous state update, a number
of technical details have to be handled. Time becomes a crucial
entity --- all variables now become, semantically, functions of
real time (and not, as in Event-B, functions of an index into
a sequence of state values). We have to have a policy about how
wild, or well behaved, the functions of time are permitted to
be, and for that, we demand that these functions have well defined
left and right limits everywhere, and are continuous from the right.
We also want to use differential equations (ODEs) to specify behaviour,
so we need to restrict to functions which are piecewise solutions to ODE
systems.

Fig.~\ref{fig-heb-mch} shows a simple Hybrid Event-B machine. 
After the machine name is the TIME declaration, which names
the variable used to denote real time (if needed). This permits
read-only access to time in the rest of the machine. Time is
synchronised (via a WHEN clause) with the start of a run in the
$\mathit{INITIALISATION}$. Next comes a CLOCK variable $clk$;
these can be reset to some value (but are otherwise
read-only) and allow time to be measured from convenient
starting points. Then come the PLIANT and VARIABLES declarations.
The former introduces the pliant variables, while the latter
introduces the mode variables. Next come the INVARIANTS. These
are required to hold {\it at all times}, so, for example the
invariant `$x \in \mathbb{R}$' means that the function of time
that is the semantic value of the variable $x$ is a {\it real
valued} function of time. A little thought shows that this is
the same convention used in Event-B (aside from time being real
rather than an index). Accepting that $x \in \mathbb{R}$ holds
at all times, does not help in knowing how values of $x$ at `nearby' 
times are related. To keep things under control, we must insist
that the semantic value of $x$ is, again, at least piecewise, a
solution to some ODE system.

The events are next. The mode events (STATUS~ordinary) are {\it exactly}
as in Event-B, and the syntax of $\mathit{MoEv}$ in Fig.~\ref{fig-heb-mch}
shows which variables can be updated by a mode event, using the generic
$\mathit{BApred}$ syntax. The pliant events (STATUS~pliant), such as
$\mathit{PliEv}$, contain novel features and require more care. The WHERE
guard is as in Event-B, imposing enabledness conditions that involve mode
variables only. The INIT guard imposes enabledness conditions that involve
the pliant variables (but which can also involve mode variables, if needed).
The actual state update is specified in the COMPLY and SOLVE clauses. Since
the purpose of a pliant event is to specify state update over an extended
period of time, some tricky technical issues arise, which we mostly ignore
in this overview. See PaperI for a full discussion.

The SOLVE clause is the easiest to describe. It can contain one
or more ODEs (with syntax $\mathscr{D}x = \phi$) and one or more
direct assignments (with syntax $y$~:=~$E$). Provided the RHSs of
these forms are as well behaved as stipulated above, no special
problems arise. Thus, direct update to something which is already well
behaved poses no problems, and integrating an ODE with well behaved
RHS improves the already acceptable behaviour as regards smoothness.
By contrast, the COMPLY clause imposes invariant-like constraints that
must hold during the execution of the pliant event, and are indicated by
$\mathit{BDApred}$, the before-during-after clause (the analogue of
Event-B's $\mathit{BApred}$). Essentially, the same caveats regarding
time dependent behaviours that we mentioned in the context of invariants
must hold for the properties that comprise $\mathit{BDApred}$.

As for almost all formal semantics of hybrid system notations,
the semantics of Hybrid Event-B is operational. Starting with the
$\mathit{INITIALISATION}$ mode event, mode and pliant event
executions alternate within a run. Each mode event execution must
enable at least one pliant event and disable all mode events --- after
which a pliant event execution then takes over. Each pliant event execution
runs until one of three things happens. (1) Some mode event becomes enabled.
At that moment the pliant event execution is preempted, and then some
enabled mode event execution takes over. (2) The running pliant event
becomes infeasible and the run stops (finite termination). (3) Neither
of the preceding options occurs and the pliant event runs forever
(nontermination). Note the language in the immediately preceding remarks.
They indicate that there are choices to be made at mode$\rightarrow$pliant
handover and pliant$\rightarrow$mode handover. These choices are
nondeterministic among all the events that are enabled at the requisite
moment. They must all be of the right kind (either all mode or all pliant),
or else the run aborts.

In more concrete terms, a run of a Hybrid Event-B machine looks like
the following. Time $\mathcal{T}$ corresponds with a semi-infinite
interval of the reals, e.g.~$\mathcal{T} = \mathbb{R}_{\geq 0}$.
The operational semantics partitions this into a sequence of
non-empty, left-closed, right-open intervals
$\mathcal{T} \equiv \langle [t_0 \ldots t_1), [t_1 \ldots t_2), \ldots
\rangle$. Thus, if $t_0$ is the initial time point, the state of the
machine's variables at $t_0$ is as specified in the $\mathit{INITIALISATION}$
mode event. During $[t_0 \ldots t_1)$, i.e.~during the set of times
$t_0 \leq t < t_1$, the state evolves as specified in the pliant event
chosen to execute immediately after $\mathit{INITIALISATION}$. This
runs until it is preempted by some mode event which is enabled at time
$t_1$. At that moment, the mode event executes, and the state changes
discontinuously. The mode event's before-state is the left limit of
the state evolution at $t_1$, and the mode event's after-state is the
value at $t_1$ itself. This pattern repeats. So $[t_1 \ldots t_2)$ is
the duration of the next pliant event execution, after which a mode
event is executed at $t_2$, followed by the next pliant event execution.
And so on.

To ensure that all executions of a Hybrid Event-B machine take place as
described, a (considerable) number of proof obligations (POs) need to be
shown to hold. Some of these, particularly concerning mode events, closely
resemble their Event-B counterparts. Others, particularly concerning pliant
events, contain novel elements. In all cases though, some potentially
delicate technicalities connected with continuous behaviour intrude into
the formulation of the POs. We do not have space to discuss these here,
so we refer to PaperI and PaperIII for details.

\section{Hybrid Event-B Refinement}
\label{sec-heb-ref}

As for all dialects of the B-Method, refinement is an important topic for
Hybrid Event-B. Given that Hybrid Event-B pertains to a context in which
real world time plays a significant role, there are a number of ways to
formulate a refinement notion. Given further that in refinement there has
to be an abstract machine and a concrete machine, the most urgent question
concerns how the notions of `real world time' are related to one another
in the two machines. We could allow the two notions to be a bit `elastic'
with respect to one another, using something like a Skorokhod metric to
measure how far apart the elasticity had stretched abstract and concrete
time and behaviour. However, this was felt to be technically too obscure
for a formalism like Hybrid Event-B, that was aimed at practical engineering
purposes. So, for Hybrid Event-B, a more conservative approach was taken,
and it became a matter of principle, that {\bf Hybrid Event-B preserves
the notion of real world time through refinement}. In fact, it can be
said that, although it is never stated, this is an assumption that also
typically applies to Event-B itself, insofar as Event-B models are intended
to reflect phenomena in the real world. If we add to this principle the
desire that Hybrid Event-B refinement does not disturb the essential
structural features of Event-B refinement, which amounts to saying that
Hybrid Event-B refinement of mode events works as does refinement of events
in Event-B, a quite strong set of constraints is generated.

\begin{figure}
\centering
\includegraphics[width=0.5\textwidth,
                 keepaspectratio=true,viewport=60 480 390 740,
                 clip=true]
{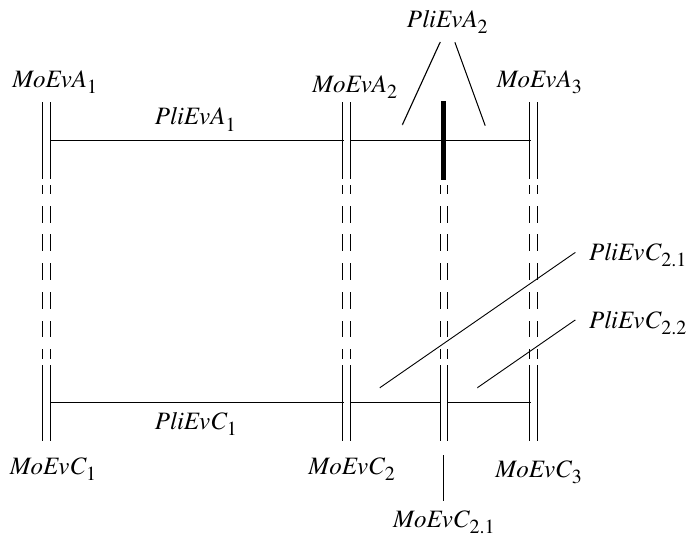}
\caption{Illustrating Hybrid Event-B refinement.}
\label{fig-heb-ref}
\end{figure}

Fig.~\ref{fig-heb-ref} shows what happens. The closely spaced vertical
bars represent the before- and after-states of mode event executions,
while the horizontal lines between them represent the continuous state
change of pliant event executions. For the moment let us forget
the occurrences of $\mathit{PliEv}$\pmb{\_} and pretend that the horizontal
lines just represent the passage of time, i.e.~that we are dealing with
Event-B refinement in a situation in which real world time passes between
event executions. The traditional
Event-B refinement proof obligations tie together what happens at abstract
and concrete levels quite closely. Thus for every abstract event execution,
there is a corresponding concrete event execution of its refinement (by
Event-B relative deadlock freedom). Conversely, for every event execution
of the concrete refinement of an abstract event, there is a corresponding
abstract event execution of its abstraction (by Event-B guard strengthening).
Between execution occurrences of abstract event refinements, execution
occurrences of `new' events, freshly introduced at the concrete level,
can appear. These must refine abstract {\sf\small skip}s (represented by
a solid vertical bar in Fig.~\ref{fig-heb-ref}).

Transferring this verbatim to the Hybrid Event-B world says how refinement
of mode events works. Fig.~\ref{fig-heb-ref} thus shows abstract
$\mathit{MoEvA}_1$, $\mathit{MoEvA}_2$, $\mathit{MoEvA}_3$, which are
refined to $\mathit{MoEvC}_1$, $\mathit{MoEvC}_2$, $\mathit{MoEvC}_3$,
at the same moments of time as their abstract counterparts. `New' mode
event $\mathit{MoEvC}_{2.1}$ executes at some point between
$\mathit{MoEvC}_2$ and $\mathit{MoEvC}_3$.

Pliant events must fit round this. An immediate consequence is that the
durations of abstract pliant event executions are equal to, or are
partitioned by, suitable concrete pliant event execution durations ---
the intervals of abstract pliant event executions are related to the
intervals of concrete pliant event executions via an inverse function.
Beyond this, the natural extension of the principle that abstract mode
events are refined by corresponding concrete mode events implies that
abstract pliant events are refined by corresponding concrete pliant
events too. In Fig.~\ref{fig-heb-ref} this would imply that
$\mathit{PliEvA}_1$ was refined by $\mathit{PliEvC}_1$ and that
$\mathit{PliEvA}_2$ was refined by $\mathit{PliEvC}_{2.1}$.

Without `new' concrete events, there would be little more to be said,
and new mode events cause little trouble by themselves. It turns out that
provided the concrete refinements of abstract events are not constrained
to have the same name, there is no possible distinction between old and
new pliant events. Observe that the duration of $\mathit{PliEvC}_{2.1}$
in Fig.~\ref{fig-heb-ref} is strictly shorter than the duration of
$\mathit{PliEvA}_2$. To close the gap in time at the concrete level,
some pliant event, $\mathit{PliEvC}_{2.2}$ must execute. There is no
alternative to the fact that $\mathit{PliEvC}_{2.2}$ must refine
$\mathit{PliEvA}_2$ --- both events have an extended duration so there
is no counterpart of the `refining of {\sf\small skip}' that can apply
to mode events. The main attendant issue that arises is the fact that
prior to the execution of $\mathit{MoEvC}_{2.1}$, $\mathit{PliEvA}_2$
was executing. And while it was not able to change any mode variables
during this period, it certainly was able to change some pliant variables,
this being its main purpose. So, by the time the guards of
$\mathit{PliEvC}_{2.2}$ need to be checked, the abstract state is, in
general, not the same as it was when the execution of $\mathit{PliEvA}_2$
was launched. This, despite the fact that the only event that the
guards of $\mathit{PliEvC}_{2.2}$ might be checked against is still
$\mathit{PliEvA}_2$. The relevant PO resolves this by making the checking
of the $iv$ guard of the corresponding abstract event optional. Checking
of the abstract $grd$ guard works, since $grd$ only involves mode variables,
which will not have changed during the time since $\mathit{PliEvA}_2$ was
launched. Whether it makes sense or not to also check the abstract $iv$
guard is highly dependent on the details of $\mathit{PliEvA}_2$ and of
the rest of the abstract machine.

As for machines, a large number of proof obligations need to be discharged
to ensure that a refinement behaves as just described. Again, we do not have
space to discuss these here, so we refer to PaperI and PaperIII for details.

\section{Multiple Hybrid Event-B Machines}
\label{sec-heb-mult}

In today's engineering landscape, having a design and development
methodology that does not cater for separate development of components,
is almost inconceivable. Therefore, since Hybrid Event-B is certainly
intended for realistic formal system development, the implications of
separate development, in the context of all the other constraints
that the formalism imposes, must be confronted. Earlier work on
combining and decomposing Event-B machines has yielded a number of
schemes, based on shared variables, shared events and interfaces
\cite{Abrial-EB:05,HoAb:10,HaHo:12,Butler:09,SilBu:09,Silv..:11}.
The Hybrid Event-B scheme, investigated in detail in PaperII, is based on
ideas from all of these, taking into account the special needs of both mode
and pliant events.

When multiple syntactic constructs have to come together to make a bigger
whole, a key issue is syntactic visibility: which elements of which entities
are visible/readable/writable in which other entities, and what restrictions
are in place to control this? This realisation was the key concept behind the
introduction of the INTERFACE construct in Hybrid Event-B. An INTERFACE is a
container that can hold a number of variables, the invariants that they must
satisfy, and their initialisations. It is thus rather like a machine without
events (or with one unstated default event simply demanding compliance with
the invariants). A machine can access an interface via a CONNECTS clause or
a READS clause. This allows the machine's events to inspect (READS) and update
(CONNECTS) the interface's variables. The interface's invariants are aggregated
with the machine's during verification. Since more than one machine can do
this, we have a mechanism for sharing variables.

In multi-machine Hybrid Event-B, to prevent a verification free-for-all that
would impede separate working, invariants are restricted to be of two kinds.
There are type I invariants (tIi's) which only mention variables from the
construct (MACHINE or INTERFACE) that declares them. By themselves these are
too restrictive to allow sufficiently expressive multi-machine working, so
there are also type II invariants (tIIi's) which are exclusive to interfaces,
and are of the form $U(u) \Rightarrow V(v)$, where $u$ are variables that
belong to one interface (the tIIi's local variables and interface) and $v$
are variables that belong to a different interface (the tIIi's remote
variables and interface). Type II invariants are declared in the interface
containing the local variables. The remote interface of a tIIi is mentioned
in the local interface using a READS declaration while the local interface
of a tIIi is mentioned in the remote interface using a REFERS declaration.

\begin{figure}
\centering
\includegraphics[width=0.6\linewidth,
                 keepaspectratio=true,viewport=20 0 560 250,
                 clip=true]
{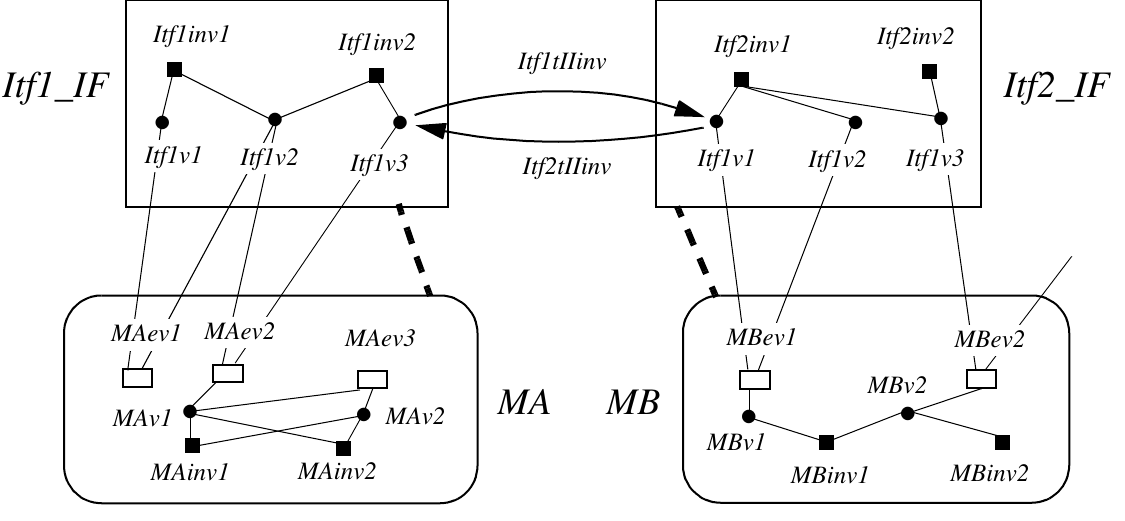}

\caption{Multiple machines and their interfaces.}
\label{fig-multi}
\end{figure}

We give a small illustration of these principles in Fig.~\ref{fig-multi}.
Small black disks represent variables, while small black squares represent
tIi's. Small rectangles represent events. Events and invariants are connected
to the variables that mention them by thin lines. Interfaces are large
rectangles containing the variables and invariants they encapsulate --- there
are two in Fig.~\ref{fig-multi}, $\mathit{Itf1\_IF}$ and $\mathit{Itf2\_IF}$.
Machines are large rounded rectangles containing their local variables,
tIi's and events --- again there are two, $\mathit{MA}$ and $\mathit{MB}$
in the figure. The very short thin line with one end free from an event in
$\mathit{MB}$, is an I/O variable connected to the environment. The CONNECTS
relationship between a machine and an interface is depicted by a thick dashed
line. Finally, tIIi's are represented by an arrow from a variable in the local
interface to a variable in the remote interface (these containing the local
and remote variable subsets of the tIIi respectively). The construct that
aggregates all the machines and interfaces of a development is the PROJECT
construct, which is not indicated in Fig.~\ref{fig-multi}.

Refinement adds some potential complexity to the above. Additional constraints
are needed to continue to prevent a verification free-for-all that would
impede separate working. It is taken as unquestionable that machines and
interfaces should both be capable of being refined. The restriction imposed
on the refinement process is that the joint invariant that couples concrete
variables to their abstract counterparts in a construct (be it a machine
or an interface), should involve only the variables declared in the two
constructs, and cannot involve variables declared in any other interface
that either construct has access to. It is clear that such a restriction
ensures that invariants proved at a high level continue to hold, suitably
translated into concrete variables through the joint invariant, at lower
levels of abstraction. 

In a single Hybrid Event-B machine, mode events and pliant events alternate.
Within a single machine context, once the nuances of continuous/discrete
behaviour are appreciated, it is fairly easy to work out how POs should be
designed that can adequately police the alternation. Things are more
complicated in a multi-machine world. We design separate machines on the
basis that their activities are expected to be largely (though obviously
not totally) independent. What does the alternation between mode events
and pliant events amount to in such an environment? The answer requires a
design decision about the multi-machine world.

One guiding principle is that every machine, at all times, is executing
some event (whether mode or pliant). Another is that in the purely discrete
Event-B world, synchronised execution of events in multiple machines is a
useful feature in some of the formulations cited above. Accordingly, the
decision was taken to permit the synchronised execution of mode events
via the introduction of a SYNCHronisation construct that declares that
families of mode events across different machines of the same project
are to be scheduled together. Of course the conjunction of the relevant
guards has to permit this to be the case.

Synchronised execution of events across multiple machines is of most
interest in the context of {\it decomposition}, in which refinement of
a single abstract event has made it too unwieldy for it to be considered
as a single indivisible entity any more, and the execution of different
parts of it in different machines makes more sense. Decomposition feels
like the converse of composition (which is what the multi-machine mechanisms
are addressing) but it is not --- the event scheduling strategies of (Hybrid)
Event-B get in the way. Thus suppose abstract machine $M$ has two mode events
$\mathit{EvA}$ and $\mathit{EvB}$, both simultaneously enabled. Then, the
scheduling strategy of $M$ prevents the simultaneous execution of
$\mathit{EvA}$ and $\mathit{EvB}$. Now suppose $\mathit{EvA}$ is decomposed
into $\mathit{EvAX}$ and $\mathit{EvAY}$, and $\mathit{EvB}$ is decomposed
into $\mathit{EvBX}$ and $\mathit{EvBY}$, and the $X$ parts execute in
decomposed machine $\mathit{MX}$, and the $Y$ parts execute in decomposed
machine $\mathit{MY}$. It is not inconceivable that in a world where the
two machines schedule their events independently, $\mathit{EvAX}$ could be
scheduled simultaneously with $\mathit{EvBY}$, giving the wrong semantics.
It is for that reason that the SYNCH mechanism was introduced in Hybrid
Event-B (following analogous mechanisms in Event-B). Decomposition of
events requires further thought regarding inputs, outputs and local
parameters. The design in PaperII requires that any input or output of
a decomposed event is confined to one of the components (rather than
being itself a synchronised activity). Local parameters are likewise
assigned by one of the components and then shared with the rest, in a
conceptually atomic single-writer multiple-readers rendezvous.

Note that we have said nothing of synchronising or decomposing pliant
events. Pliant events raise their own issues in a multi-machine world.
We recall that pliant events executing in a machine typically get preempted. 
If there are multiple machines, the preemptions in different machines
should be independent, otherwise the different machines are behaving
like a single machine. This presents little problem in the real world
context, but creates technical complexity for a formal operational
semantics which is concerned with constructing a consistent global
system trace. At a preemption in machine $\mathit{MA}$, the non-preempted
other machines must be `paused' and their execution resumed after the
scheduling choices in machine $\mathit{MA}$ have been resolved. And
because this is a strange thing to contemplate for the physical world,
the physical world consequences of any such technicalities must be
invisible. This affects potential decomposition strategies for pliant
events. In the light of these considerations, the strategy adopted
for decomposition of pliant events is that it has to be programmed
explicitly by the user. If pliant events are required to execute
simultaneously in two machines, their guards must be arranged to
be the same (and the same as the guards of the parent event if they
constitute a decomposition), and to prevent `the wrong parts' of
different decompositions from being scheduled simultaneously, no
parent events in the same machine that are to be decomposed can
ever be enabled simultaneously. Fortunately, it is easy enough
to impose static restrictions on event guards that can ensure
the conditions required. A full discussion can be found in PaperII.

\section{A Small Incident Response Case Study}
\label{sec-inc-resp}

We illustrate the utility of the multi-machine Hybrid Event-B approach
with a small case study. For lack of space we do not cover all aspects
in equal depth, but focus on those structural elements which particularly
contribute to convenient exploration of distributed and autonomous
systems.

The case study concerns an incident response system. This consists of
a number of largely independent agents: a controller, three responders
that enter a potentially hazardous area to effect some appropriate
measures, and two drones, that act as communication intermediaries
between the controller (which is assumed to be ground based) and the
responders (which are assumed to be unable to communicate with the
controller directly). Since the arena of the incident needing the response
is assumed to be hazardous, two drones are provided, and they are expected
to keep apart from each other, in case one of them suffers some disabling
mishap. Described in such abstract terms, the case study is applicable to
situations from natural and industrial disasters, through terrorist attacks,
to battlefield warfighting.

Each of these agents is modelled using an individual Hybrid Event-B machine,
and there are additionally CONTEXTs and INTERFACEs that organise the means
by which the various agents can cooperate. The ability to do this successfully
results from the ability to partition the functionality needed into constructs
that capture convincingly self-contained subsets, and the adequate flexibility
of the cooperation mechanisms, that enables their shared goals to be achieved.
These mechanisms have to cope with six independent but cooperating agents
in the complete system, each potentially involving independently determined
smooth state change.

The essential lesson of the case study is that the design of multi-machine
Hybrid Event-B permits the system to be described in such a way that
experimentation with different scenarios can be easily achieved by changing
just one machine and one context. These are $\bf EnvironmentScenario\_Mch$
and $\bf IncidentResponse\_CTX$.

In the body of this paper we focus on the machines that capture the
behaviours of the agents, relegating the contexts and interfaces
that contain the static declarations that support these machines to
the Appendix. Here and there we take some liberties with Event-B syntax
for the sake of readability. Next, we see
$\bf EnvironmentScenario\_Mch$.

\vspace{1.5ex}
{\scriptsize 
\noindent
\fbox{%
\parbox[t]{0.48\linewidth}{
\hspace*{2ex}MACHINE~~$\bf EnvironmentScenario\_Mch$\rule{0ex}{2.5ex}\\
\hspace*{2ex}SEES~~$\mathit{IncidentResponse\_CTX}$\\
\hspace*{2ex}CONNECTS~~$\mathit{IncidentResponse\_IF}$\\
\hspace*{2ex}VARIABLES\\
\hspace*{4ex}$schedule$\\
\hspace*{2ex}INVARIANTS\\
\hspace*{4ex}$schedule : \mathrm{seq}(\mathbb{R})$\\
\hspace*{4ex}$\mathrm{increasing}(schedule)$\\
\hspace*{2ex}EVENTS\\
\hspace*{4ex}$\mathit{INITIALISATION}$\\
\hspace*{6ex}STATUS~~ordinary\\
\hspace*{6ex}BEGIN\\
\hspace*{8ex}$schedule$ ~:=~ $\mathit{INITSCHED}$\\
\hspace*{6ex}END\\
\hspace*{4ex}$\mathit{PliTrue}$\\
\hspace*{6ex}STATUS~~pliant\\
\hspace*{6ex}COMPLY\\
\hspace*{8ex}$\mathit{INVARIANTS}$\\
\hspace*{6ex}END\\
\hspace*{2ex}\ldots~~~~\ldots
\raisebox{-1ex}{\rule{0pt}{2.5ex}}
}}
\hfill
\fbox{%
\parbox[t]{0.48\linewidth}{
\hspace*{2ex}\ldots~~~~\ldots\rule{0ex}{2.5ex}\\
\hspace*{4ex}$\mathit{AddHazard}$\\
\hspace*{6ex}STATUS~~ordinary\\
\hspace*{6ex}ANY~~$tg,xx,yy,sz,ht$\\
\hspace*{6ex}WHERE\\
\hspace*{8ex}$\textrm{nonempty}(schedule)~\land$\\
\hspace*{8ex}$t = \mathrm{head}(schedule)~\land$\\
\hspace*{8ex}$(tg \mapsto xx \mapsto yy \mapsto sz \mapsto ht) \notin hazards$\\
\hspace*{6ex}BEGIN\\
\hspace*{8ex}$hazards$ ~:=~ $hazards \cup
             \{tg \mapsto xx \mapsto yy \mapsto sz \mapsto ht\}$ \\
\hspace*{8ex}$schedule$ ~:=~ $\textrm{tail}(schedule)$\\
\hspace*{6ex}END\\
\hspace*{4ex}$\mathit{TakeHazard}$\\
\hspace*{6ex}STATUS~~ordinary\\
\hspace*{6ex}ANY~~$tg,xx,yy,sz,ht$\\
\hspace*{6ex}WHERE\\
\hspace*{8ex}$\textrm{nonempty}(schedule)~\land$\\
\hspace*{8ex}$t = \mathrm{head}(schedule)~\land$\\
\hspace*{8ex}$(tg \mapsto xx \mapsto yy \mapsto sz \mapsto ht) \in hazards$\\
\hspace*{6ex}BEGIN\\
\hspace*{8ex}$hazards$ ~:=~ $hazards -
             \{tg \mapsto xx \mapsto yy \mapsto sz \mapsto ht\}$ \\
\hspace*{8ex}$schedule$ ~:=~ $\textrm{tail}(schedule)$\\
\hspace*{6ex}END\\
\hspace*{2ex}END
\raisebox{-1ex}{\rule{0pt}{2.5ex}}
}}
}

\vspace{1.5ex}
\noindent
The $\bf EnvironmentScenario\_Mch$ machine above contains a $schedule$
of interventions into the incident arena. These are restricted to be the
introduction and removal of hazardous areas within the arena, and the
$schedule$ variable is a succession (i.e.~sequence) of times at which the
interventions take place. Each hazardous area has either a $SQ$uare footprint
or a $CYL$indrical shape. So each is specified using a $SQ$ or $CYL$ tag $tg$,
$x$ and $y$ coordinates, a size $sz$, and a height $ht$. The
$\bf EnvironmentScenario\_Mch$ machine uses the $\mathit{AddHazard}$ event
to introduce a fresh hazard when time reaches the first value in $schedule$,
with the attributes of the hazard, written in B-Method notation as
$(tg \mapsto xx \mapsto yy \mapsto sz \mapsto ht)$, being taken from the
formal model's environment using the ANY clause. The $\mathit{TakeHazard}$
event removes an existing hazard from the set of current hazards $hazards$.
The $schedule$ itself is statically defined in the $\bf IncidentResponse\_CTX$
context. Changing the constants in the context, and the details of the
$\bf EnvironmentScenario\_Mch$ machine is all one needs to do to
experiment with different incident management scenarios in this world.

Below, we see the project file, $\bf IncidentResponse\_Prj$. Its purpose
is to orchestrate all the components of the system. Many specific details
are curtailed to save space. The first declared item is the GLOBalINVariantS
file $\bf IncidentResponse\_GI$ which we discuss at the end.
After that there is a list of contexts, interfaces and machines.
The line `CONTEXT $\bf Drone1\_CTX$ IS' instantiates the context as
a version of a generic library context for drones, $\bf Drone\_CTX$.
After the WITH, there is a renaming mapping saying how identifiers in
$\bf Drone\_CTX$ should be replaced to get $\bf Drone1\_CTX$. This
instantiation allows the two contexts $\bf Drone1\_CTX$ and $\bf Drone2\_CTX$
to have different constants $V\!dr1$ and $V\!dr2$ for the velocities of the
drones $\bf Drone1\_Mch$ and $\bf Drone2\_Mch$ that SEE the respective
contexts. After the interfaces there are the machines.
$\bf Controller\_Mch$ is discussed below, while $\bf EnvironmentScenario\_Mch$
was discussed above. After these, there are the two drone instantiations
$\bf Drone1\_Mch$ and $\bf Drone2\_Mch$, being instantiations of a generic
$\bf Drone\_Mch$. We have suppressed the details of the instantiations,
which amount to adding `1' or `2' to identifiers in the generic machine.

Beyond instantiations, the project file contains SYNCH lines. These specify
collections of mode events across multiple machines that must be scheduled
simultaneously (i.e.~only when all their guards are simultaneously true).
Thus `SYNCH $\bf ActivateDrone1$' enforces the simultaneous execution
of event

\vspace{1ex}
{\scriptsize 
\noindent
\fbox{%
\parbox[t]{0.48\linewidth}{
\hspace*{2ex}PROJECT~~$\bf IncidentResponse\_Prj$\rule{0ex}{2.5ex}\\
\hspace*{2ex}GLOBINVS~~$\bf IncidentResponse\_GI$\\
\hspace*{2ex}CONTEXT~~$\bf IncidentResponse\_CTX$\\
\hspace*{2ex}CONTEXT~~$\bf Controller\_CTX$\\
\hspace*{2ex}CONTEXT~~$\bf Drone1\_CTX$~~IS\\
\hspace*{4ex}$\mathit{Drone\_CTX}$~~WITH\\
\hspace*{6ex}$Vdr \rightarrow Vdr1$\\
\hspace*{4ex}END\\
\hspace*{2ex}CONTEXT~~$\bf Drone2\_CTX$~~IS\\
\hspace*{4ex}$\mathit{Drone\_CTX}$~~WITH\\
\hspace*{6ex}$Vdr \rightarrow Vdr2$\\
\hspace*{4ex}END\\
\hspace*{2ex}CONTEXT~~$\bf Responder\_CTX$\\
\hspace*{2ex}INTERFACE~~$\bf IncidentResponse\_IF$\\
\hspace*{2ex}INTERFACE~~$\bf ControllerDrones\_IF$\\
\hspace*{2ex}INTERFACE~~$\bf ControllerResponder\_IF$\\
\hspace*{2ex}MACHINE~~$\bf Controller\_Mch$\\
\hspace*{2ex}MACHINE~~$\bf EnvironmentScenario\_Mch$\\
\hspace*{2ex}MACHINE~~$\bf Drone1\_Mch$~~IS\\
\hspace*{4ex}$\mathit{Drone\_Mch}$~~WITH\\
\hspace*{6ex}$\bullet\bullet\bullet$\\
\hspace*{4ex}END\\
\hspace*{2ex}MACHINE~~$\bf Drone2\_Mch$~~IS\\
\hspace*{4ex}$\mathit{Drone\_Mch}$~~WITH\\
\hspace*{6ex}$\bullet\bullet\bullet$\\
\hspace*{4ex}END\\
\hspace*{2ex}MACHINE~~$\bf Responder1\_Mch$~~IS\\
\hspace*{4ex}$\mathit{Responder\_Mch}$~~WITH\\
\hspace*{6ex}$\bullet\bullet\bullet$\\
\hspace*{4ex}END\\
\hspace*{2ex}MACHINE~~$\bf Responder2\_Mch$~~IS\\
\hspace*{4ex}$\mathit{Responder\_Mch}$~~WITH\\
\hspace*{6ex}$\bullet\bullet\bullet$\\
\hspace*{4ex}END\\
\hspace*{2ex}\ldots~~~~\ldots
\raisebox{-1ex}{\rule{0pt}{2.5ex}}
}}
\hfill
\fbox{%
\parbox[t]{0.48\linewidth}{
\hspace*{2ex}\ldots~~~~\ldots\rule{0ex}{2.5ex}\\
\hspace*{2ex}MACHINE~~$\bf Responder3\_Mch$~~IS\\
\hspace*{4ex}$\mathit{Responder\_Mch}$~~WITH\\
\hspace*{6ex}$\bullet\bullet\bullet$\\
\hspace*{4ex}END\\
\hspace*{2ex}SYNCH $\bf ActivateDrone1$\\
\hspace*{6ex}$\mathit{Controller\_Mch.LaunchDrone1}$\\
\hspace*{6ex}$\mathit{Drone1\_Mch.Activate1}$\\
\hspace*{4ex}END\\
\hspace*{2ex}SYNCH $\bf UpdateDrone1$\\
\hspace*{6ex}$\mathit{Controller\_Mch.UpdateDrone1}$\\
\hspace*{6ex}$\mathit{Drone1\_Mch.Update1}$\\
\hspace*{4ex}END\\
\hspace*{2ex}SYNCH $\bf DeActivateDrone1$\\
\hspace*{6ex}$\mathit{Controller\_Mch.RecallDrone1}$\\
\hspace*{6ex}$\mathit{Drone1\_Mch.DeActivate1}$\\
\hspace*{4ex}END\\
\hspace*{2ex}SYNCH~~~$\bullet\bullet\bullet~~~\bf Drone2$\\
\hspace*{2ex}SYNCH $\bf ActivateResponder1$\\
\hspace*{6ex}$\mathit{Controller\_Mch.LaunchResponder1}$\\
\hspace*{6ex}$\mathit{Responder1\_Mch.Activate1}$\\
\hspace*{4ex}END\\
\hspace*{2ex}SYNCH $\bf UpdateResponder1$\\
\hspace*{6ex}$\mathit{Controller\_Mch.UpdateResponder1}$\\
\hspace*{6ex}$\mathit{Responder1\_Mch.Update1}$\\
\hspace*{4ex}END\\
\hspace*{2ex}SYNCH $\bf DeActivateResponder1$\\
\hspace*{6ex}$\mathit{Controller\_Mch.RecallResponder1}$\\
\hspace*{6ex}$\mathit{Responder1\_Mch.DeActivate1)}$\\
\hspace*{4ex}END\\
\hspace*{2ex}SYNCH~~~$\bullet\bullet\bullet~~~\bf Responder2$\\
\hspace*{2ex}SYNCH~~~$\bullet\bullet\bullet~~~\bf Responder3$\\
\hspace*{2ex}END
\raisebox{-1ex}{\rule{0pt}{2.5ex}}
}}
}

\noindent
$\bf LaunchDrone1$ in the $\bf Controller\_Mch$ machine and of
$\bf Activate1$ in the $\bf Drone1\_Mch$ machine. There are similar
synchronisations for $\bf Upate$ing and for $\bf DeActivate$ing the
$\bf Drone1\_Mch$ machine. The analogous synchronisations for the
$\bf Drone2\_Mch$ machine are suppressed. A similar pattern applies to
the responders. The details for the $\bf Responder1\_Mch$ machine are
given in full, while those for machines\linebreak
$\bf Responder2\_Mch$ and $\bf Responder3\_Mch$ are suppressed.

{\scriptsize 
\noindent
\fbox{%
\parbox[t]{0.48\linewidth}{
\hspace*{2ex}MACHINE~~$\bf Controller\_Mch$\rule{0ex}{2.5ex}\\
\hspace*{2ex}SEES~~$\bf IncidentResponse\_CTX$\\
\hspace*{2ex}SEES~~$\bf Controller\_CTX$\\
\hspace*{2ex}CONNECTS~~$\bf IncidentResponse\_IF$\\
\hspace*{2ex}CONNECTS~~$\bf ControllerDrone1\_IF$\\
\hspace*{2ex}CONNECTS~~$\bf ControllerDrone2\_IF$\\
\hspace*{2ex}CONNECTS~~$\bf ControllerResponder1\_IF$\\
\hspace*{2ex}CONNECTS~~$\bf ControllerResponder2\_IF$\\
\hspace*{2ex}CONNECTS~~$\bf ControllerResponder3\_IF$\\
\hspace*{2ex}VARIABLES\\
\hspace*{4ex}$mode$\\
\hspace*{4ex}$ctrhazards$\\
\hspace*{4ex}$cyclestart$\\
\hspace*{4ex}$drones2comd$\\
\hspace*{4ex}$responders2comd$\\
\hspace*{2ex}INVARIANTS\\
\hspace*{4ex}$mode : \mathit{CTRLSTATE}$\\
\hspace*{4ex}$ctrhazards : \mathbb{P}(\mathit{HAZTYPE} \times
   \mathbb{R} \times \mathbb{R} \times \mathbb{R} \times \mathbb{R})$\\
\hspace*{4ex}$cyclestart : \mathbb{R}$\\
\hspace*{4ex}$drones2comd : \pow(\{1,2\})$\\
\hspace*{4ex}$responders2comd : \pow(\{1,2,3\})$\\
\hspace*{2ex}EVENTS\\
\hspace*{4ex}$\mathit{INITIALISATION}$\\
\hspace*{6ex}STATUS~~ordinary\\
\hspace*{6ex}BEGIN\\
\hspace*{8ex}$mode$ ~:=~ $\mathit{OFF}$\\
\hspace*{8ex}$ctrhazards$ ~:=~ $\emptyset$\\
\hspace*{8ex}$cyclestart$ ~:=~ $0$\\
\hspace*{8ex}$drones2comd$ ~:=~ $\emptyset$\\
\hspace*{8ex}$resps2comd$ ~:=~ $\emptyset$\\
\hspace*{6ex}END\\
\hspace*{4ex}$\mathit{PliTrue}$\\
\hspace*{6ex}STATUS~~pliant\\
\hspace*{6ex}COMPLY\\
\hspace*{8ex}$\mathit{INVARIANTS}$\\
\hspace*{6ex}END\\
\hspace*{4ex}$\mathit{ActivateController}$\\
\hspace*{6ex}STATUS~~asynch\\
\hspace*{6ex}WHEN\\
\hspace*{8ex}$mode = \mathit{OFF} \land 0 < t < \delta$\\
\hspace*{6ex}THEN\\
\hspace*{8ex}$mode$ ~:=~ $\mathit{DISPATCH}$\\
\hspace*{8ex}$drones2comd$ ~:=~ $\{1,2\}$\\
\hspace*{8ex}$resps2comd$ ~:=~ $\{1,2,3\}$\\
\hspace*{6ex}END\\
\hspace*{4ex}$\mathit{LaunchDrone1}$\\
\hspace*{6ex}STATUS~~asynch\\
\hspace*{6ex}WHEN\\
\hspace*{8ex}$mode = \mathit{DISPATCH} \land 1 \in drones2comd \land t < \delta$\\
\hspace*{6ex}THEN\\
\hspace*{8ex}$drhazards$ ~:=~ $hazards$\\
\hspace*{8ex}$drones2comd$ ~:=~ $drones2comd - \{1\}$\\
\hspace*{6ex}END\\
\hspace*{4ex}$\mathit{LaunchDrone2}$\\
\hspace*{8ex}$\bullet\bullet\bullet$\\
\hspace*{6ex}END\\
\hspace*{2ex}\ldots~~~~\ldots
\raisebox{-1ex}{\rule{0pt}{2.5ex}}
}}
\hfill
\fbox{%
\parbox[t]{0.48\linewidth}{
\hspace*{2ex}\ldots~~~~\ldots\rule{0ex}{2.5ex}\\
\hspace*{4ex}$\mathit{LaunchResp1}$\\
\hspace*{6ex}STATUS~~asynch\\
\hspace*{6ex}WHEN\\
\hspace*{8ex}$mode = \mathit{DISPATCH}~\land$\\
\hspace*{8ex}$1 \in resps2comd \land t < \delta$\\
\hspace*{6ex}THEN\\
\hspace*{8ex}$resp1hazards$ ~:=~ $hazards$\\
\hspace*{8ex}$resps2comd$ ~:=~ $resps2comd - \{1\}$\\
\hspace*{6ex}END\\
\hspace*{4ex}$\mathit{LaunchResp2}$\\
\hspace*{8ex}$\bullet\bullet\bullet$\\
\hspace*{6ex}END\\
\hspace*{4ex}$\mathit{LaunchResp3}$\\
\hspace*{8ex}$\bullet\bullet\bullet$\\
\hspace*{6ex}END\\
\hspace*{4ex}$\mathit{StartMonitoring}$\\
\hspace*{6ex}STATUS~~asynch\\
\hspace*{6ex}WHEN\\
\hspace*{8ex}$mode = \mathit{DISPATCH} \land t = \delta$\\
\hspace*{6ex}THEN\\
\hspace*{8ex}$mode$ ~:=~ $\mathit{UPDATEHAZ}$\\
\hspace*{8ex}$cyclestart$ ~:=~ $\Delta$\\
\hspace*{6ex}END\\
\hspace*{4ex}$\mathit{MonitorHazardsNull}$\\
\hspace*{6ex}STATUS~~asynch\\
\hspace*{6ex}WHEN\\
\hspace*{8ex}$mode = \mathit{UPDATEHAZ} \land 0 < t - cyclestart < \delta/2~\land$\\
\hspace*{8ex}$hazards = ctrhazards$\\
\hspace*{6ex}THEN\\
\hspace*{8ex}$cyclestart$ ~:=~ $cyclestart + \Delta$\\
\hspace*{6ex}END\\
\hspace*{4ex}$\mathit{MonitorHazardsUpdate}$\\
\hspace*{6ex}STATUS~~asynch\\
\hspace*{6ex}WHEN\\
\hspace*{8ex}$mode = \mathit{UPDATEHAZ} \land 0 < t - cyclestart < \delta/2~\land$\\
\hspace*{8ex}$hazards \neq ctrhazards$\\
\hspace*{6ex}THEN\\
\hspace*{8ex}$ctrhazards$ ~:=~ $hazards$\\
\hspace*{8ex}$drones2comd$ ~:=~ $\{1,2\}$\\
\hspace*{8ex}$resps2comd$ ~:=~ $\{1,2,3\}$\\
\hspace*{8ex}$cyclestart$ ~:=~ $cyclestart + \Delta$\\
\hspace*{6ex}END\\
\hspace*{4ex}$\mathit{UpdateDrone1}$\\
\hspace*{6ex}STATUS~~asynch\\
\hspace*{6ex}WHEN\\
\hspace*{8ex}$mode = \mathit{UPDATEHAZ} \land 1 \in drones2comd~\land$\\
\hspace*{10ex}$cyclestart - \Delta + \delta/2 < t < cyclestart - \Delta + \delta$\\
\hspace*{6ex}THEN\\
\hspace*{8ex}$drhazards$ ~:=~ $hazards$\\
\hspace*{8ex}$drones2comd$ ~:=~ $drones2comd - \{1\}$\\
\hspace*{6ex}END\\
\hspace*{4ex}$\mathit{UpdateDrone2}$\\
\hspace*{8ex}$\bullet\bullet\bullet$\\
\hspace*{6ex}END\\
\hspace*{2ex}\ldots~~~~\ldots
\raisebox{-1ex}{\rule{0pt}{2.5ex}}
}}
}

{\scriptsize 
\noindent
\fbox{%
\parbox[t]{0.48\linewidth}{
\hspace*{2ex}\ldots~~~~\ldots\rule{0ex}{2.5ex}\\
\hspace*{4ex}$\mathit{UpdateResp1}$\\
\hspace*{6ex}STATUS~~asynch\\
\hspace*{6ex}WHEN\\
\hspace*{8ex}$mode = \mathit{UPDATEHAZ} \land 1 \in resps2comd~\land$\\
\hspace*{10ex}$cyclestart - \Delta + \delta/2 < t < cyclestart - \Delta + \delta$\\
\hspace*{6ex}THEN\\
\hspace*{8ex}$resp1hazards$ ~:=~ $hazards$\\
\hspace*{8ex}$resps2comd$ ~:=~ $resps2comd - \{1\}$\\
\hspace*{6ex}END\\
\hspace*{4ex}$\mathit{UpdateResp2}$\\
\hspace*{8ex}$\bullet\bullet\bullet$\\
\hspace*{6ex}END\\
\hspace*{4ex}$\mathit{UpdateResp3}$\\
\hspace*{8ex}$\bullet\bullet\bullet$\\
\hspace*{6ex}END\\
\hspace*{4ex}$\mathit{EndMonitoring}$\\
\hspace*{6ex}STATUS~~asynch\\
\hspace*{6ex}WHEN\\
\hspace*{8ex}$mode = \mathit{UPDATEHAZ}~\land$\\
\hspace*{10ex}$\mathit{DURATION} < t < \mathit{DURATION} + \delta$\\
\hspace*{6ex}THEN\\
\hspace*{8ex}$mode$ ~:=~ $\mathit{RECALL}$\\
\hspace*{8ex}$drones2comd$ ~:=~ $\{1,2\}$\\
\hspace*{8ex}$resps2comd$ ~:=~ $\{1,2,3\}$\\
\hspace*{6ex}END\\
\hspace*{2ex}\ldots~~~~\ldots
\raisebox{-1ex}{\rule{0pt}{2.5ex}}
}}
\hfill
\fbox{%
\parbox[t]{0.48\linewidth}{
\hspace*{2ex}\ldots~~~~\ldots\rule{0ex}{2.5ex}\\
\hspace*{4ex}$\mathit{RecallDrone1}$\\
\hspace*{6ex}STATUS~~asynch\\
\hspace*{6ex}WHEN\\
\hspace*{8ex}$mode = \mathit{RECALL} \land 1 \in drones2comd~\land$\\
\hspace*{10ex}$\mathit{DURATION} < t < \mathit{DURATION} + \delta$\\
\hspace*{6ex}THEN\\
\hspace*{8ex}$drones2comd$ ~:=~ $drones2comd - \{1\}$\\
\hspace*{6ex}END\\
\hspace*{4ex}$\mathit{RecallDrone2}$\\
\hspace*{8ex}$\bullet\bullet\bullet$\\
\hspace*{6ex}END\\
\hspace*{4ex}$\mathit{RecallResp1}$\\
\hspace*{6ex}STATUS~~asynch\\
\hspace*{6ex}WHEN\\
\hspace*{8ex}$mode = \mathit{RECALL} \land 1 \in resps2comd~\land$\\
\hspace*{10ex}$\mathit{DURATION} < t < \mathit{DURATION} + \delta$\\
\hspace*{6ex}THEN\\
\hspace*{8ex}$resps2comd$ ~:=~ $resps2comd - \{1\}$\\
\hspace*{6ex}END\\
\hspace*{4ex}$\mathit{RecallResp2}$\\
\hspace*{8ex}$\bullet\bullet\bullet$\\
\hspace*{6ex}END\\
\hspace*{4ex}$\mathit{RecallResp3}$\\
\hspace*{8ex}$\bullet\bullet\bullet$\\
\hspace*{6ex}END\\
\hspace*{4ex}$\mathit{DeActivateController}$\\
\hspace*{6ex}STATUS~~asynch\\
\hspace*{6ex}WHEN\\
\hspace*{8ex}$mode = \mathit{RECALL}~\land$\\
\hspace*{10ex}$\mathit{DURATION} + \delta < t < \mathit{DURATION} + 2\delta$\\
\hspace*{6ex}THEN\\
\hspace*{8ex}$mode$ ~:=~ $\mathit{OFF}$\\
\hspace*{6ex}END\\
\hspace*{2ex}END\raisebox{-1ex}{\rule{0pt}{2.5ex}}
}}
}

\vspace{1.5ex}
Following the project file, there is the most complex machine of the
project, the $\bf Controller\_Mch$ machine, occupying the four
panels above. Essentially, it is a finite state machine, perfectly
expressible using Event-B alone if need be, except that the availability
of real time for scheduling purposes simplifies the state space that
would otherwise be needed.

The main job of the controller, once activated (event
$\mathit{ActivateController}$) is: to send forth the drones and
responders (events $\mathit{LaunchDrone1}$, $\mathit{LaunchDrone2}$,
$\mathit{LaunchResp1}$, $\mathit{LaunchResp2}$, $\mathit{LaunchResp3}$),
to poll the environment machine to discover changes to the collection of
hazards that have been unearthed (events $\mathit{StartMonitoring}$,
$\mathit{MonitorHazardsNull}$, $\mathit{MonitorHazardsUpdate}$) and to
advise the drones and responders of the same (events $\mathit{UpdateDrone1}$,
$\mathit{UpdateDrone2}$, $\mathit{UpdateResp1}$, $\mathit{UpdateResp2}$,
$\mathit{UpdateResp3}$), and finally to recall the drones and
responders when the work has been completed (events $\mathit{RecallDrone1}$,
$\mathit{RecallDrone2}$, $\mathit{RecallResp1}$, $\mathit{RecallResp2}$,
$\mathit{RecallResp3}$). Once this is done, the controller is deactivated
(event $\mathit{DeActivateController}$). A default $\mathit{PliTrue}$
pliant event covers the gaps between state change.

In the above, there are clearly groups of similar events that need to take
place periodically (launching, monitoring, updating, recalling) for agents
that are largely independent of each other most of the time. Achieving this
using a pure finite state machine would entail either a state explosion to
accommodate all possible interleavings of individual agent events, or an
excessive use of sequentialisation to avoid it. The presence of time as an
intrinsic feature in Hybrid Event-B (assumed synchronised across all machines)
permits a more economical approach.

A `large' time window $\Delta$ is defined along with a `small' window
$\delta$. Launching is defined to take place within the first $\delta$
of elapsed time, via synchronised events, as described above. Since
the main synchronisation mechanism is the time interval of duration
$\delta$, arbitrary orderings of the launching events within that
interval are permitted.

The same technique is used for all the other synchronised activities. 
Monitoring and update take place repeatedly after the elapse of each
period of duration $\Delta$. At $t = n\,\Delta$, the
EnvironmentScenario machine potentially updates the $hazards$.
Within the next $\delta/2$ period, the controller checks whether $hazards$
has changed since the last check, and if so, schedules updates to the
drones and responders to update their hazard data and to replan if
necessary. To keep the model simple, this is accomplished by setting the
$drones2comd$ and $resps2comd$ variables to the IDs of all the drones
and responders. This triggers, in the next $\delta/2$ period, each of
them to update it's local copy of the $hazards$ variable via a synchronised
event such as e.g.,~the combination of $\mathit{UpdateDrone1}$ in the
controller together with $\mathit{Update}$ in first drone. Once that has
happened, the drone or responder can recalculate its trajectory.

\veryslightlylongpage
\vspace{1ex}
{\scriptsize 
\noindent
\fbox{%
\parbox[t]{0.48\linewidth}{
\hspace*{2ex}MACHINE~~$\bf Drone\_Mch$\rule{0ex}{2.5ex}\\
\hspace*{2ex}SEES~~$\bf IncidentResponse\_CTX$\\
\hspace*{2ex}SEES~~$\bf Drone\_CTX$\\
\hspace*{2ex}CONNECTS~~$\bf ControllerDrones\_IF$\\
\hspace*{2ex}VARIABLES\\
\hspace*{4ex}$mode$\\
\hspace*{4ex}$thex~,~they~,~thez$\\
\hspace*{4ex}$trajectory$\\
\hspace*{2ex}INVARIANTS\\
\hspace*{4ex}$mode : \mathit{DRONESTATE}$\\
\hspace*{4ex}$thex~,~they~,~thez : \mathbb{R}~,~\mathbb{R}~,~\mathbb{R}$\\
\hspace*{4ex}$trajectory :
    \mathrm{seq}(\mathbb{R} \times \mathbb{R} \times \mathbb{R})$\\
\hspace*{2ex}EVENTS\\
\hspace*{4ex}$\mathit{INITIALISATION}$\\
\hspace*{6ex}STATUS~~ordinary\\
\hspace*{6ex}BEGIN\\
\hspace*{8ex}$mode$ ~:=~ $\mathit{OFF}$\\
\hspace*{8ex}$thex~,~they~,~thez$ ~:=~ $0~,~0~,~0$\\
\hspace*{8ex}$trajectory$ ~:=~ $\langle\,\rangle$\\
\hspace*{6ex}END\\
\hspace*{4ex}$\mathit{PliTrue}$\\
\hspace*{6ex}STATUS~~pliant\\
\hspace*{6ex}WHEN\\
\hspace*{8ex}$mode = \mathit{OFF}$\\
\hspace*{6ex}COMPLY\\
\hspace*{8ex}$\mathit{INVARIANTS}$\\
\hspace*{6ex}END\\
\hspace*{4ex}$\mathit{Activate}$\\
\hspace*{6ex}STATUS~~ordinary\\
\hspace*{6ex}WHEN\\
\hspace*{8ex}$mode = \mathit{OFF}$\\
\hspace*{6ex}BEGIN\\
\hspace*{8ex}$mode$ ~:=~ $\mathit{SEEK}$\\
\hspace*{8ex}$thex~,~they~,~thez$ ~:=~ $drx~,~dry~,~drz$\\
\hspace*{8ex}$trajectory$ ~:=~ $calcCentAvoidTraj(\ldots)$\\
\hspace*{6ex}END\\
\hspace*{4ex}$\mathit{Update}$\\
\hspace*{6ex}STATUS~~ordinary\\
\hspace*{6ex}WHEN\\
\hspace*{8ex}$mode \in \{\mathit{SEEK}~,~\mathit{RETURN}\}$\\
\hspace*{6ex}BEGIN\\
\hspace*{8ex}$thex~,~they~,~thez$ ~:=~ $drx~,~dry~,~drz$\\
\hspace*{8ex}$trajectory$ ~:=~ $calcCentAvoidTraj(\ldots)$\\
\hspace*{6ex}END\\
\hspace*{2ex}\ldots~~~~\ldots
\raisebox{-1ex}{\rule{0pt}{2.5ex}}
}}
\hfill
\fbox{%
\parbox[t]{0.48\linewidth}{
\hspace*{2ex}\ldots~~~~\ldots\rule{0ex}{2.5ex}\\
\hspace*{4ex}$\mathit{Navigate}$\\
\hspace*{6ex}STATUS~~pliant\\
\hspace*{6ex}WHEN\\
\hspace*{8ex}$mode \in \{\mathit{SEEK}~,~\mathit{RETURN}\}~\land$\\
\hspace*{8ex}$trajectory \neq \langle\,\rangle$\\
\hspace*{6ex}SOLVE\\
\hspace*{8ex}$\mathscr{D}drx$ ~:=~ $V\!dr \times
   (\mathrm{first}(trajectory)[1] - thex)$\\
\hspace*{8ex}$\mathscr{D}dry$ ~:=~ $V\!dr \times
   (\mathrm{first}(trajectory)[2] - they)$\\
\hspace*{8ex}$\mathscr{D}drz$ ~:=~ $V\!dr \times
   (\mathrm{first}(trajectory)[3] - thez)$\\
\hspace*{6ex}END\\
\hspace*{4ex}$\mathit{Waypoint}$\\
\hspace*{6ex}STATUS~~ordinary\\
\hspace*{6ex}WHEN\\
\hspace*{8ex}$mode \in \{\mathit{SEEK}~,~\mathit{RETURN}\}~\land$\\
\hspace*{8ex}$drx = \mathrm{first}(trajectory)[1]~\land$\\
\hspace*{8ex}$dry = \mathrm{first}(trajectory)[2]~\land$\\
\hspace*{8ex}$drz = \mathrm{first}(trajectory)[3]~\land$\\
\hspace*{8ex}$trajectory \neq \langle\,\rangle$\\
\hspace*{6ex}BEGIN\\
\hspace*{8ex}$thex~,~they~,~thez$ ~:=~ $drx~,~dry~,~drz$\\
\hspace*{8ex}$trajectory$ ~:=~ $\mathrm{rest}(trajectory)$\\
\hspace*{6ex}END\\
\hspace*{4ex}$\mathit{Hover}$\\
\hspace*{6ex}STATUS~~pliant\\
\hspace*{6ex}WHEN\\
\hspace*{8ex}$mode \in \{\mathit{SEEK}~,~\mathit{RETURN}\}~\land$\\
\hspace*{8ex}$trajectory = \langle\,\rangle$\\
\hspace*{6ex}COMPLY\\
\hspace*{8ex}$\mathit{INVARIANTS}$\\
\hspace*{6ex}END\\
\hspace*{4ex}$\mathit{DeActivate}$\\
\hspace*{6ex}STATUS~~ordinary\\
\hspace*{6ex}WHEN\\
\hspace*{8ex}$mode = \mathit{SEEK}$\\
\hspace*{6ex}BEGIN\\
\hspace*{8ex}$mode$ ~:=~ $\mathit{RETURN}$\\
\hspace*{8ex}$thex~,~they~,~thez$ ~:=~ $drx~,~dry~,~drz$\\
\hspace*{8ex}$trajectory$ ~:=~ $calcCentAvoidTraj(\ldots)$\\
\hspace*{6ex}END\\
\hspace*{4ex}$\mathit{SwitchOff}$\\
\hspace*{6ex}STATUS~~ordinary\\
\hspace*{6ex}WHEN\\
\hspace*{8ex}$mode = \mathit{RETURN}~\land$\\
\hspace*{8ex}$drx = dry = drz = 0$\\
\hspace*{6ex}BEGIN\\
\hspace*{8ex}$mode$ ~:=~ $\mathit{OFF}$\\
\hspace*{6ex}END\\
\hspace*{2ex}END\raisebox{-1ex}{\rule{0pt}{2.5ex}}
}}
}

The cycle of updating and recalculating continues until it is time to
recall all the drones and responders. This is achieved by the same
synchonisation mechanism. In other words, after $\mathit{DURATION}$
has elapsed, similar windows of length $\delta$ are established and
the drones and responders adopt first the $\mathit{RETURN}$ mode,
and after returning home, the $\mathit{OFF}$ mode.

The generic drone machine $\bf Drone\_Mch$ appears in the two panels on
the previous page. Once a drone is activated through the synchronised
$\mathit{Activate}$ event, it ceases the default $\mathit{PliTrue}$
behaviour and instead pursues the $\mathit{Navigate}$ behaviour.
The preceding mode event occurrence, whether an $\mathit{Activate}$
or an $\mathit{Update}$ event, caused it to remember its 3D position
at that time in variables $thex$, $they$, $thez$, from\linebreak

\vspace{-1.5ex}
{\scriptsize 
\noindent
\fbox{%
\parbox[t]{0.48\linewidth}{
\hspace*{2ex}MACHINE~~$\bf Responder\_Mch$\rule{0ex}{2.5ex}\\
\hspace*{2ex}SEES~~$\bf Responder\_CTX$\\
\hspace*{2ex}CONNECTS~~$\bf ControllerResponder\_IF$\\
\hspace*{2ex}VARIABLES\\
\hspace*{4ex}$mode$\\
\hspace*{4ex}$thex~,~they$\\
\hspace*{4ex}$trajectory$\\
\hspace*{2ex}PLIANT\\
\hspace*{4ex}$respx~,~respy$\\
\hspace*{2ex}INVARIANTS\\
\hspace*{4ex}$mode : \mathit{RESPSTATE}$\\
\hspace*{4ex}$thex~,~they : \mathbb{R}~,~\mathbb{R}$\\
\hspace*{4ex}$respx~,~respy : \mathbb{R}~,~\mathbb{R}$\\
\hspace*{4ex}$trajectory :
    \mathrm{seq}(\mathbb{R} \times \mathbb{R})$\\
\hspace*{2ex}EVENTS\\
\hspace*{4ex}$\mathit{INITIALISATION}$\\
\hspace*{6ex}STATUS~~ordinary\\
\hspace*{6ex}BEGIN\\
\hspace*{8ex}$mode$ ~:=~ $\mathit{OFF}$\\
\hspace*{8ex}$thex~,~they$ ~:=~ $0~,~0$\\
\hspace*{8ex}$respx~,~respy$ ~:=~ $0~,~0$\\
\hspace*{8ex}$trajectory$ ~:=~ $\langle\,\rangle$\\
\hspace*{6ex}END\\
\hspace*{4ex}$\mathit{PliTrue}$\\
\hspace*{6ex}STATUS~~pliant\\
\hspace*{6ex}WHEN\\
\hspace*{8ex}$mode = \mathit{OFF}$\\
\hspace*{6ex}COMPLY\\
\hspace*{8ex}$\mathit{INVARIANTS}$\\
\hspace*{6ex}END\\
\hspace*{4ex}$\mathit{Activate}$\\
\hspace*{6ex}STATUS~~ordinary\\
\hspace*{6ex}WHEN\\
\hspace*{8ex}$mode = \mathit{OFF}$\\
\hspace*{6ex}BEGIN\\
\hspace*{8ex}$mode$ ~:=~ $\mathit{SEEK}$\\
\hspace*{8ex}$thex~,~they$ ~:=~ $respx~,~respy$\\
\hspace*{8ex}$trajectory$ ~:=~ $calcTraj(\ldots)$\\
\hspace*{6ex}END\\
\hspace*{4ex}$\mathit{Update}$\\
\hspace*{6ex}STATUS~~ordinary\\
\hspace*{6ex}WHEN\\
\hspace*{8ex}$mode \in \{\mathit{SEEK}~,~\mathit{RETURN}\}$\\
\hspace*{6ex}BEGIN\\
\hspace*{8ex}$thex~,~they$ ~:=~ $respx~,~respy$\\
\hspace*{8ex}$trajectory$ ~:=~ $calcTraj(\ldots)$\\
\hspace*{6ex}END\\
\hspace*{2ex}\ldots~~~~\ldots
\raisebox{-1ex}{\rule{0pt}{2.5ex}}
}}
\hfill
\noindent
\fbox{%
\parbox[t]{0.48\linewidth}{
\hspace*{2ex}\ldots~~~~\ldots\rule{0ex}{2.5ex}\\
\hspace*{4ex}$\mathit{Navigate}$\\
\hspace*{6ex}STATUS~~pliant\\
\hspace*{6ex}WHEN\\
\hspace*{8ex}$mode \in \{\mathit{SEEK}~,~\mathit{RETURN}\}~\land$\\
\hspace*{8ex}$trajectory \neq \langle\,\rangle$\\
\hspace*{6ex}SOLVE\\
\hspace*{8ex}$\mathscr{D}respx$ ~:=~ $V\!dr \times
   (\mathrm{first}(trajectory)[1] - thex)$\\
\hspace*{8ex}$\mathscr{D}respy$ ~:=~ $V\!dr \times
   (\mathrm{first}(trajectory)[2] - they)$\\
\hspace*{6ex}END\\
\hspace*{4ex}$\mathit{Waypoint}$\\
\hspace*{6ex}STATUS~~ordinary\\
\hspace*{6ex}WHEN\\
\hspace*{8ex}$mode \in \{\mathit{SEEK}~,~\mathit{RETURN}\}~\land$\\
\hspace*{8ex}$respx = \mathrm{first}(trajectory)[1]~\land$\\
\hspace*{8ex}$respy = \mathrm{first}(trajectory)[2]~\land$\\
\hspace*{8ex}$trajectory \neq \langle\,\rangle$\\
\hspace*{6ex}BEGIN\\
\hspace*{8ex}$thex~,~they$ ~:=~ $respx~,~respy$\\
\hspace*{8ex}$trajectory$ ~:=~ $\mathrm{rest}(trajectory)$\\
\hspace*{6ex}END\\
\hspace*{4ex}$\mathit{Arrived}$\\
\hspace*{6ex}STATUS~~asynch\\
\hspace*{6ex}WHEN\\
\hspace*{8ex}$mode = SEEK \land trajectory = \langle\,\rangle$\\
\hspace*{6ex}BEGIN\\
\hspace*{8ex}$mode$ ~:=~ $ARRIVED$\\
\hspace*{6ex}END\\
\hspace*{4ex}$\mathit{DoSomethingForAWhile}$\\
\hspace*{6ex}STATUS~~pliant\\
\hspace*{6ex}WHEN\\
\hspace*{8ex}$mode = ARRIVED$\\
\hspace*{6ex}COMPLY\\
\hspace*{8ex}$\mathit{INVARIANTS}$\\
\hspace*{6ex}END\\
\hspace*{4ex}$\mathit{DeActivate}$\\
\hspace*{6ex}STATUS~~ordinary\\
\hspace*{6ex}WHEN\\
\hspace*{8ex}$mode = \mathit{ARRIVED}$\\
\hspace*{6ex}BEGIN\\
\hspace*{8ex}$mode$ ~:=~ $\mathit{RETURN}$\\
\hspace*{8ex}$thex~,~they$ ~:=~ $respx~,~respy$\\
\hspace*{8ex}$trajectory$ ~:=~ $calcTraj(\ldots)$\\
\hspace*{6ex}END\\
\hspace*{4ex}$\mathit{SwitchOff}$\\
\hspace*{6ex}STATUS~~ordinary\\
\hspace*{6ex}WHEN\\
\hspace*{8ex}$mode = \mathit{RETURN}~\land$\\
\hspace*{8ex}$respx = respy = respz = 0$\\
\hspace*{6ex}BEGIN\\
\hspace*{8ex}$mode$ ~:=~ $\mathit{OFF}$\\
\hspace*{6ex}END\\
\hspace*{2ex}END
\raisebox{-1ex}{\rule{0pt}{2.5ex}}
}}
}

\noindent
which it can calculate a trajectory towards its goal. The tactic taken is
to navigate towards the centroid of the positions of the controller and
responders. The two drones need to ensure that they avoid damage in
a hazardous area, so if they need to fly close to one, they need to
take heed of its geometry, including height. Also, to avoid simultaneous
destruction of both the drones though mishap, they need to ensure that
they stay far enough apart from each other. This is all done by the
$\mathit{calcCentAvoidTraj}(\ldots)$ function that they use, except
that, in this paper, we abstract away from the detailed calculations
that would be needed to achieve this. Evidently, the drones need to
share their positions to do all this, so they share a single interface
which declares both, which they can use to exchange positions by
communicating with each other. For reasons of brevity, in this paper
we do not cover this aspect in detail either. 

All of this missing detail could be handled via a suitable Hybrid Event-B
refinement. For simplicity again, a trajectory calculated by the drone consists
of a sequence of line segments between points in three dimensions, which
the drones follow, in order. The traversal of each segment is accomplished
within the $\mathit{Navigate}$ event. This simply increments each of the
spatial coordinates, using the drone velocity (which is statically defined
in its context) in proportion to the coordinate difference to be traversed.
Once a segment is completed, the first segment is deleted from the trajectory
in the $\mathit{Waypoint}$ event, and the drone follows the next segment,
until there is no trajectory left, at which point the drone stays where
it is, maintaining distance from its partner.

The last big machine of the development is the generic responder machine
$\bf Responder\_Mch$ which appears on the previous page. This is 
similar to the drone machine in most respects, notably the technique
used for synchronisation with the controller, but with a few essential
differences. For example, the responders stay on the ground, so do not
need a $z$ coordinate. They are also assumed to not need to communicate
with each other regarding position, so in our model, each responder can
have its own interface. They do not need to fly above hazards, so
their trajectory calculations will be different, expressed in the
function $\mathit{calcTraj}(\ldots)$. And once arrived, they do some
work until they, along with the drones, are recalled by the controller,
which is achieved using the customary mechanisms.

\section{Verification of Hybrid Event-B Models}
\label{sec-verif}

The main purpose of the Hybrid Event-B approach is the verification of
models expressed within the formalism, in order to derive increased
confidence about their dependability. As for its predecessors in the
classical B-Method and Event-B, the emphasis is on safety properties,
expressed within the invariants declared in the model's syntactic
constructs. In the case of Hybrid Event-B, these can be found inside
machines and interfaces. At the time of writing, there is no tool
support for Hybrid Event-B, so a comprehensive mechanically supported
verification exercise cannot be reported on. Nevertheless, the POs
that define correctness in Hybrid Event-B have been investigated in
detail in the papers cited earlier, so an outline of what would be
involved can be given.

Two things deserve to be highlighted. The first is that, as indicated
earlier, the design of multi-machine Hybrid Event-B is such that the
syntactic scope of all the POs is precisely defined by the rules of
construction that models have to conform to. The second is that, the
present model has deliberately been designed to be so simple, that
the overwhelming majority of POs become rather trivial.

One set of POs concerns feasibility: i.e., is there an after-state for every
enabled mode event? The simple assignments to constants in their bodies
says yes; likewise for pliant events, which have either trivial COMPLY
{\it INVARIANTS} bodies, or merely specify the following of a linear
trajectory. Another major set of POs concerns the preservation of the
invariants. Since, by design, these are almost exclusively trivial typing
declarations, the answer will again be yes. Such POs can be dealt with by
inspection. Slightly more complicated are POs that deal with correct
handover: mode/pliant and pliant/mode. This involves calculating disjunctions
of guards, to ensure that when an event completes, an event of the right kind
can succeede it. Again, extreme simplicity makes this a task that can be
done by hand. So we can be confident our model is as it should be, even if
mechanical corroboration would be even better.

The preceding remarks prepare the ground for discussing the last construct
of the project, the global invariants $\bf IncidentResponse\_GI$. The global
invariants offer the possibility of explicitly stating invariants of the system
that are properties of the system as a whole, and not merely of some part of
it. Clearly such a capability is important when safety properties depend on
the safe cooperation of all parts of the system. Global invariants are
intended to be derivable from the remaining invariants of a correct model,
but this is not a hard constraint. The intention is that global invariants
would be checked in the latter stages of a development, once the correctness
of the constituent parts of the project is established.

\vspace{1ex}
\noindent
{\scriptsize 
\noindent
\fbox{%
\parbox[t]{0.98\linewidth}{
\hspace*{2ex}GLOBINVS~~$\bf IncidentResponse\_GI$\rule{0ex}{2.5ex}\\
\hspace*{2ex}SEES~~$\bf IncidentResponse\_CTX$\\
\hspace*{2ex}CONNECTS~~$\bf IncidentResponse\_IF$\\
\hspace*{2ex}INVARIANTS\\
\hspace*{4ex}$t \notin \Union\,(ii\,\bullet\,ii \in \dom(\mathit{INITSCHED}) ~|~
    [\mathit{INITSCHED}(ii) \ldots \mathit{INITSCHED}(ii)+\delta])
    \Rightarrow$\\
\hspace*{8ex}$(~(hazards = ctrhazards) \land (hazards = drhazards)~\land$\\
\hspace*{9ex}$\,(hazards = resp1hazards)~\land (hazards = resp2hazards) \land
    (hazards = resp3hazards)~)$\\
\hspace*{2ex}END
\raisebox{-1ex}{\rule{0pt}{2.5ex}}
}}
}

\vspace{1ex}
The $\bf IncidentResponse\_GI$ construct above contains the following rather
simple property of this kind. It states that outside the periods when the
individual machines' perceptions of the hazard configurations are being
updated (which are the time intervals of length $\delta$ following those
integral multiples of $\Delta$ defined in the $\mathit{INITSCHED}$ constant
of the IncidentResponse context), they all agree on the hazards.

The conclusion of the invariant is a conjunction of statements each of
which is a property of a part of the system. So it cannot be proved before
all the pieces have been successfully constructed. A technical detail is
that, for simplicity, the individual conjuncts (appropriately guarded)
were not included as invariants of the relevant machines above. This itself
does not make the global invariant unprovable of course. Global invariants,
in particular, underline the connection between the Hybrid Event-B approach
to system correctness and the reference to safety properties in this paper's
title.

\section{Conclusions}
\label{sec-conc}

In the early part of the paper, we surveyed Hybrid Event-B and how it
evolved from Event-B. The inclusion of real time and smooth state change
entails considerable additional technical complexity in the semantics,
and we discussed the details of this to the extent that space permitted.
Multi-machine working brings in a whole raft of additional technical
details to worry about, principally concerned with synchronisations
and with the interplay of structure and verification needs. We surveyed
this as far as possible. The multi-machine Hybrid Event-B approach is
particularly suited to the formalisation and verification of autonomous
systems, as these systems are seldom isolated, must fend for themselves
for long periods of time while nevertheless communicating, but only
intermittently, and are often cyber-physical. So modelling them naturally
partitions into a constellation of cooperating but largely self-contained
Hybrid Event-B machines, supported by suitable syntactic constructs.

The point of all this was to lay the groundwork for a small multi-machine
example, to support the case just made. A simple system of intermittently
communicating drones, responders and a controller was modelled. It is
reasonable to say that this illustrated convincingly the capabilities of
multi-machine Hybrid Event-B to fluently capture the kinds of behaviour
pattern needed in such systems.

\newpage

\bibliographystyle{eptcs}
\bibliography{zzcontevb}

\newpage

\appendix

\section{Appendix: Contexts, Interfaces and Instantiated Machine Outlines}
\label{sec-app}

{\scriptsize 
\noindent
\fbox{\parbox[t]{0.48\linewidth}{
\hspace*{2ex}CONTEXT~~$\bf IncidentResponse\_CTX$\rule{0ex}{2.5ex}\\
\hspace*{2ex}CONSTANTS\\
\hspace*{4ex}$SQ~,~CYL$\\
\hspace*{4ex}$\mathit{INITSCHED}$\\
\hspace*{4ex}$\delta~,~\Delta$\\
\hspace*{4ex}$\mathit{DURATION}$\\
\hspace*{4ex}$\mathit{RESP1dest}~,~\mathit{RESP2dest}~,~\mathit{RESP3dest}$\\
\hspace*{2ex}SETS\\
\hspace*{4ex}$\mathit{HAZTYPE}$\\
\hspace*{2ex}AXIOMS\\
\hspace*{4ex}$partition(\mathit{HAZTYPE},\{SQ\},\{CYL\})$\\
\hspace*{4ex}$\delta : \mathbb{R}~,~\delta = 0.1$\\
\hspace*{4ex}$\Delta : \mathbb{R}~,~\Delta = 1$\\
\hspace*{2ex}\ldots~~~~\ldots
\raisebox{-1ex}{\rule{0pt}{2.5ex}}
}}
\hfill
\fbox{%
\parbox[t]{0.48\linewidth}{
\hspace*{2ex}\ldots~~~~\ldots\rule{0ex}{2.5ex}\\
\hspace*{4ex}$\mathit{INITSCHED} : \mathrm{seq}(\mathbb{R})$\\
\hspace*{4ex}$\mathit{INITSCHED} = \langle 12, 30, 55 \rangle$\\
\hspace*{4ex}$DURATION : \mathbb{R}$\\
\hspace*{4ex}$DURATION = 79.7$\\
\hspace*{4ex}$\mathit{RESP1dest} : \mathbb{R} \times \mathbb{R}$\\
\hspace*{4ex}$\mathit{RESP1dest} = (12.3 \mapsto 15.0)$\\
\hspace*{4ex}$\mathit{RESP2dest} : \mathbb{R} \times \mathbb{R}$\\
\hspace*{4ex}$\mathit{RESP2dest} = (-11.2 \mapsto 14.0)$\\
\hspace*{4ex}$\mathit{RESP3dest} : \mathbb{R} \times \mathbb{R}$\\
\hspace*{4ex}$\mathit{RESP3dest} = (2.1 \mapsto 29.0)$\\
\hspace*{2ex}THEOREMS\\
\hspace*{4ex}$\mathit{DURATION} > \mathrm{last}(\mathit{INITSCHED})$\\
\hspace*{2ex}END
\raisebox{-1ex}{\rule{0pt}{2.5ex}}
}}
}

\vspace{0.3ex}
{\scriptsize 
\noindent
\fbox{\parbox[t]{0.48\linewidth}{
\hspace*{2ex}INTERFACE~~$\bf IncidentResponse\_IF$\rule{0ex}{2.5ex}\\
\hspace*{2ex}SEES~~$IncidentResponse\_CTX$\\
\hspace*{2ex}TIME~~$t$\\
\hspace*{2ex}VARIABLES\\
\hspace*{4ex}$hazards$\\
\hspace*{2ex}\ldots~~~~\ldots
\raisebox{-1ex}{\rule{0pt}{2.5ex}}
}}
\hfill
\fbox{%
\parbox[t]{0.48\linewidth}{
\hspace*{2ex}\ldots~~~~\ldots\rule{0ex}{2.5ex}\\
\hspace*{2ex}INVARIANTS\\
\hspace*{4ex}$hazards : \mathbb{P}(\mathit{HAZTYPE} \times
   \mathbb{R} \times \mathbb{R} \times \mathbb{R} \times \mathbb{R})$\\
\hspace*{6ex}:--- sq/cyl,
             $(x,y)$ coords, size from $(x,y)$, height\\
\hspace*{2ex}INITIALISATION\\
\hspace*{4ex}$t$ ~:=~ $0$ \\
\hspace*{4ex}$hazards$ ~:=~ $\emptyset$ \\
\hspace*{2ex}END
\raisebox{-1ex}{\rule{0pt}{2.5ex}}
}}
}

\vspace{0.3ex}
{\scriptsize 
\noindent
\fbox{\parbox[t]{0.48\linewidth}{
\hspace*{2ex}CONTEXT~~$\bf Controller\_CTX$\rule{0ex}{2.5ex}\\
\hspace*{2ex}CONSTANTS\\
\hspace*{4ex}$\mathit{OFF}~,~\mathit{DISPATCH}~,~
\mathit{RECALL}~,~\mathit{UPDATEHAZ}$\\
\hspace*{2ex}SETS\\
\hspace*{4ex}$\mathit{CTRLSTATE}$\\
\hspace*{2ex}\ldots~~~~\ldots
\raisebox{-1ex}{\rule{0pt}{2.5ex}}
}}
\hfill
\fbox{%
\parbox[t]{0.48\linewidth}{
\hspace*{2ex}\ldots~~~~\ldots\rule{0ex}{2.5ex}\\
\hspace*{2ex}AXIOMS\\
\hspace*{4ex}$partition(\mathit{CTRLSTATE}~,~
    \{\mathit{OFF}\}~,~\{\mathit{DISPATCH}\}~,~$\\
\hspace*{6ex}$\{\mathit{UPDATEHAZ}\}~,~\{\mathit{RECALL}\})$\\
\hspace*{2ex}END
\raisebox{-1ex}{\rule{0pt}{2.5ex}}
}}
}

\vspace{0.3ex}
{\scriptsize 
\noindent
\fbox{\parbox[t]{0.48\linewidth}{
\hspace*{2ex}CONTEXT~~$\bf Drone\_CTX$\rule{0ex}{2.5ex}\\
\hspace*{2ex}CONSTANTS\\
\hspace*{4ex}$\mathit{OFF}~,~\mathit{SEEK}~,~\mathit{RETURN}$\\
\hspace*{4ex}$V\!dr$\\
\hspace*{2ex}SETS\\
\hspace*{4ex}$\mathit{DRONESTATE}$\\
\hspace*{2ex}\ldots~~~~\ldots
\raisebox{-1ex}{\rule{0pt}{2.5ex}}
}}
\hfill
\fbox{%
\parbox[t]{0.48\linewidth}{
\hspace*{2ex}\ldots~~~~\ldots\rule{0ex}{2.5ex}\\
\hspace*{2ex}AXIOMS\\
\hspace*{4ex}$partition(\mathit{DRONESTATE}~,~$\\
\hspace*{6ex}$\{\mathit{OFF}\}~,~\{\mathit{SEEK}\}~,~\{\mathit{RETURN}\})$\\
\hspace*{4ex}$Vdr : \mathbb{R}$\\
\hspace*{2ex}END
\raisebox{-1ex}{\rule{0pt}{2.5ex}}
}}
}

\vspace{0.3ex}
{\scriptsize 
\noindent
\fbox{\parbox[t]{0.48\linewidth}{
\hspace*{2ex}INTERFACE~~$\bf ControllerDrones\_IF$\rule{0ex}{2.5ex}\\
\hspace*{2ex}TIME~~$t$\\
\hspace*{2ex}VARIABLES\\
\hspace*{4ex}$drhazards$\\
\hspace*{2ex}PLIANT\\
\hspace*{4ex}$dr1x~,~dr1y~,~dr1z$\\
\hspace*{4ex}$dr2x~,~dr2y~,~dr2z$\\
\hspace*{2ex}\ldots~~~~\ldots
\raisebox{-1ex}{\rule{0pt}{2.5ex}}
}}
\hfill
\fbox{%
\parbox[t]{0.48\linewidth}{
\hspace*{2ex}\ldots~~~~\ldots\rule{0ex}{2.5ex}\\
\hspace*{2ex}INVARIANTS\\
\hspace*{4ex}$drhazards : \mathbb{P}(\mathit{HAZTYPE} \times
   \mathbb{R} \times \mathbb{R} \times \mathbb{R} \times \mathbb{R})$\\
\hspace*{4ex}$dr1x~,~dr1y~,~dr1z : \mathbb{R}~,~\mathbb{R}~,~\mathbb{R}$\\
\hspace*{4ex}$dr2x~,~dr2y~,~dr2z : \mathbb{R}~,~\mathbb{R}~,~\mathbb{R}$\\
\hspace*{2ex}INITIALISATION\\
\hspace*{4ex}$t$ ~:=~ $0$ \\
\hspace*{4ex}$drhazards$ ~:=~ $\emptyset$\\
\hspace*{4ex}$dr1x~,~dr1y~,~dr1z$ ~:=~ $0~,~0~,~0$\\
\hspace*{4ex}$dr2x~,~dr2y~,~dr2z$ ~:=~ $0~,~0~,~0$\\
\hspace*{2ex}END
\raisebox{-1ex}{\rule{0pt}{2.5ex}}
}}
}

\newpage

{\scriptsize 
\noindent
\fbox{\parbox[t]{0.48\linewidth}{
\hspace*{2ex}CONTEXT~~$\bf Responder\_CTX$\rule{0ex}{2.5ex}\\
\hspace*{2ex}CONSTANTS\\
\hspace*{4ex}$\mathit{OFF}~,~\mathit{SEEK}~,~\mathit{ARRIVED}~,~\mathit{RETURN}$\\
\hspace*{2ex}SETS\\
\hspace*{4ex}$\mathit{RESPSTATE}$\\
\hspace*{2ex}AXIOMS\\
\hspace*{4ex}$partition(\mathit{RESPSTATE}~,~\{\mathit{OFF}\}~,~$\\
\hspace*{6ex}$\{\mathit{SEEK}\}~,~\{\mathit{ARRIVED}\}~,~\{\mathit{RETURN}\})$\\
\hspace*{2ex}END
\raisebox{-1ex}{\rule{0pt}{2.5ex}}
}}
\hfill
\noindent
\fbox{\parbox[t]{0.48\linewidth}{
\hspace*{2ex}INTERFACE~~$\bf ControllerResponder\_IF$\rule{0ex}{2.5ex}\\
\hspace*{2ex}TIME~~$t$\\
\hspace*{2ex}VARIABLES\\
\hspace*{4ex}$resphazards$\\
\hspace*{2ex}INVARIANTS\\
\hspace*{4ex}$resphazards : \mathbb{P}(\mathit{HAZTYPE} \times
   \mathbb{R} \times \mathbb{R} \times \mathbb{R} \times \mathbb{R})$\\
\hspace*{2ex}INITIALISATION\\
\hspace*{4ex}$t$ ~:=~ $0$ \\
\hspace*{4ex}$resphazards$ ~:=~ $\emptyset$ \\
\hspace*{2ex}END
\raisebox{-1ex}{\rule{0pt}{2.5ex}}
}}
}

\vspace{0.3ex}
{\scriptsize 
\noindent
\fbox{\parbox[t]{0.48\linewidth}{
\hspace*{2ex}MACHINE~~$\bf Drone1\_Mch$\rule{0ex}{2.5ex}\\
\hspace*{4ex}$\bullet\bullet\bullet$\\
\hspace*{2ex}END\\
\hspace*{2ex}MACHINE~~$\bf Drone2\_Mch$\\
\hspace*{4ex}$\bullet\bullet\bullet$\\
\hspace*{2ex}END\\
\hspace*{2ex}\ldots~~~~\ldots
\raisebox{-1ex}{\rule{0pt}{2.5ex}}
}}
\hfill
\noindent
\fbox{%
\parbox[t]{0.48\linewidth}{
\hspace*{2ex}\ldots~~~~\ldots\rule{0ex}{2.5ex}\\
\hspace*{2ex}MACHINE~~$\bf Responder1\_Mch$\\
\hspace*{4ex}$\bullet\bullet\bullet$\\
\hspace*{2ex}END\\
\hspace*{2ex}MACHINE~~$\bf Responder2\_Mch$\\
\hspace*{4ex}$\bullet\bullet\bullet$\\
\hspace*{2ex}END\\
\hspace*{2ex}MACHINE~~$\bf Responder3\_Mch$\\
\hspace*{4ex}$\bullet\bullet\bullet$\\
\hspace*{2ex}END
\raisebox{-1ex}{\rule{0pt}{2.5ex}}
}}
}

\end{document}